\documentclass[sigconf]{acmart} 

\usepackage{subcaption}
\copyrightyear{2021} 
\acmYear{2021} 
\setcopyright{licensedothergov}\acmConference[EASE 2021]{Evaluation and Assessment in Software Engineering}{June 21--23, 2021}{Trondheim, Norway}
\acmBooktitle{Evaluation and Assessment in Software Engineering (EASE 2021), June 21--23, 2021, Trondheim, Norway}
\acmPrice{15.00}
\acmDOI{10.1145/3463274.3463805}
\acmISBN{978-1-4503-9053-8/21/06}

\begin{document}
\title{Towards offensive language detection and reduction in four Software Engineering communities}

\author{Jithin Cheriyan}
\affiliation{%
  \institution{Department of Information Science\\University of Otago}
  \streetaddress{University of Otago}
  \city{Dunedin}
  \country{New Zealand}}
  \email{jithin.cheriyan@postgrad.otago.ac.nz}

\author{Bastin Tony Roy Savarimuthu}
\affiliation{%
  \institution{Department of Information Science\\University of Otago}
  \streetaddress{University of Otago}
  \city{Dunedin}
  \country{New Zealand}}
  \email{tony.savarimuthu@otago.ac.nz}

\author{Stephen Cranefield}
\affiliation{%
  \institution{Department of Information Science\\University of Otago}
  \streetaddress{University of Otago}
  \city{Dunedin}
  \country{New Zealand}}
  \email{stephen.cranefield@otago.ac.nz}

\begin{abstract}
  Software Engineering (SE) communities such as Stack Overflow have become unwelcoming, particularly through members' use of offensive language. Research has shown that offensive language drives users away from active engagement within these platforms. This work aims to explore this issue more broadly by investigating the nature of offensive language in comments posted by users in four prominent SE platforms – GitHub, Gitter, Slack and Stack Overflow (SO). It proposes an approach to detect and classify offensive language in SE communities by adopting natural language processing and deep learning techniques. Further, a Conflict Reduction System (CRS), which identifies offence and then suggests what changes could be made to minimize offence has been proposed. Beyond showing the prevalence of offensive language in over 1 million comments from four different communities which ranges from 0.07\% to 0.43\%, our results show promise in successful detection and classification of such language. The CRS system has the potential to drastically reduce manual moderation efforts to detect and reduce offence in SE communities.
\end{abstract}



\maketitle
\section{Introduction}
Any communication that contains gutter language, swearing or racist terms or content that may be considered as offensive on moral, social, religious or cultural grounds can be defined as offensive \citep{xu2010filtering}. Use of offensive or abusive language is a longstanding issue in all leading social media platforms \citep{hern_facebook_2016}. It has been noted that online SE communities are also not free from this predicament \citep{burnap2015cyber, kumar2017army}. Though the panacea for this unwelcoming practice is yet to be devised, major SE platforms like SO have started to work towards reducing offensive language. Communities such as  GitHub and SO have community moderation system, i.e., selected community members monitor, flag and sometimes delete the offensive contents, which to some extent reduces offensive language. However, human efforts to curtail offensive language in online platforms are highly time-consuming because of the volume, and the process may overlook some unpleasant comments that enter the discussion space. A sample of such deleted comments, collected from the sample public data dump of SO \citep{noauthor_can_nodate}, is given below\footnote{Reader discretion is advised owing to the nature of the work.}.   

\vspace{-7pt}
\begin{figure}[ht]
    \setlength\abovecaptionskip{-5px}
    \includegraphics[width=\linewidth, height=1.9cm]{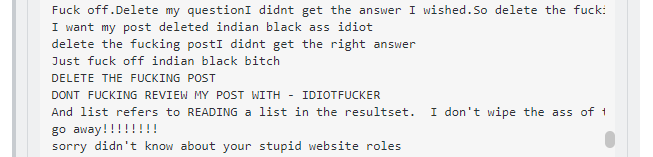}
    \setlength\belowcaptionskip{-12px}
    \caption{Sample offensive comments from SO}
    \label{Sample SO Offensive comments}
\end{figure}

These denigrating comments shatter the decorum of the sites, create conflict among members and may affect the mental and emotional well-being of members \citep{gren_importance_2018}. Any unproductive interpersonal dispute or discourse that causes friction or frustration among community members can be perceived as conflict \citep{gren_importance_2018}. Moreover, some members (particularly newcomers) tend to leave online communities that are unwelcoming or offensive \citep{culture_stack_2018}. For these reasons, automatic detection and prevention of offensive contents entering SE discussion space needs attention.

To detect offensive contents, other than human moderators, some sites employ Machine Learning (ML) techniques. Heat Detection bot\footnote{\url{https://github.com/SOBotics/HeatDetector}} used by SO is an example where the bot automatically detects offensive contents using a set of ML and Natural Language Processing (NLP) tools. If a post crosses the borders of civility, the bot flags the post and notifies the moderators. Moderators assess the post and if the flagging appears to be legitimate, they delete the post. The deletion process is similar when multiple members (humans) flag a post as offensive. The post gets deleted if the flag count exceeds a particular threshold. Therefore, though the moderation system in these platforms is laborious for humans, a human in the loop is inevitable for the sophisticated working of the system like pardoning context-specific offensive comments (i.e., allowing these comments to remain based on human judgement). However, we posit that human effort can be reduced by devising better ML approaches.

Though GitHub and SO follow community moderation involving humans, the two other communities that we consider in this work (i.e., Gitter and Slack) do not pursue the same. Slack utilises an offence detection bot\footnote{\url{https://github.com/foqal/slack-moderator}} to monitor and delete offensive contents without human involvement; however Gitter does not employ any of these strategies.

Although plenty of works have investigated abuse in social networking sites like Twitter and Facebook \citep{Koushik_2019,DBLP:conf/itasec/VignaCDPT17}, abuse in SE communities is seldom addressed. \citet{Raman_2020} have investigated the presence of toxicity in GitHub comments and the performance of various offensive content filtering tools was evaluated by \citet{Sarker_CORR}. However, no prior work has investigated the prevalence of offensive language causing conflict among community members across SE communities. \citet{gren_importance_2018} have shown that conflict between values (i.e., what is expected does not match what is observed such as non-offensive language) causes severe impact on SE development teams, which must be handled accordingly to save resources such as time and productivity.

We propose a novel system called CRS, which detects and classifies offensive comments, signals the writer about the offensive contents and finally proposes a set of paraphrased statements to reduce potential conflict owing to abuse. Therefore, the research questions that we address in this work are:

{\bfseries RQ1: What is the prevalence of offensive language across four SE communities \textendash{} GitHub, Gitter, Slack and SO, and what classes do they belong to?

RQ2: How can offensive language be reduced in SE communities?
}

The paper is structured as follows. Section ~\ref{Related works}  discusses the related works in offensive language detection and conflict reduction in SE communities. Section ~\ref{RQ1} presents the methodology and the results for the detection and classification of offensive language in SE communities, thus answering RQ1. Section ~\ref{RQ2} proposes the CRS framework to reduce conflict owing to offensive language in SE communities, thus providing one possible solution to RQ2. Section ~\ref{Discussion} discusses the implications of the results and concludes the paper.

\section{Related work}
\label{Related works}
\subsection{Offensive language detection}
Online abuse has been a pressing issue for decades owing to the widespread use of the internet, which forces nations to enact legislation to curtail hate speech on online platforms \citep{hern_facebook_2016}. Staring from the earliest work of \citet{spertus1997smokey}, many researchers investigated online abuse in various social networking sites. Various classes of  abuse detection include targeted \citep{DBLP:conf/icwsm/SilvaMCBW16,DBLP:conf/icwsm/ElSheriefKNWB18} and multi-modal hate speech detection \citep{DBLP:conf/wacv/GomezGGK20}, to cite a few. 

One of the main reasons for hate speech is the anonymity \citep{burnap2015cyber} of user posts. Owing to anonymity, people tend to become more deindividuated to express freely \citep{Festinger_1952}. Even though offenders can be blocked, certain users continue to re-offend  \citep{burnap2015cyber}. Re-offending and sock-puppeting \citep{kumar2017army} are also key reasons for offensive language prevalence in online communities.

Offensive behaviour of users keeps members away from online platforms\footnote{\url{https://sites.google.com/view/alw3/}}. Particularly in SE platforms, use of offensive or abusive terms makes these platforms less inclusive since courtesy is expected\footnote{\label{footnote2}\url{https://stackoverflow.com/conduct}}. SO officials in a blog article have admitted that the site had become unwelcoming to newcomers owing to this hostile nature \citep{culture_stack_2018}. They reported that nearly 7\% of the total comments in SO are unwelcoming, and less than 1\% of the total comments are abusive, and abusive contents are 10\% more frequent in question-related comments compared to answer-related ones \citep{silge_welcome_2018}.

Works on abuse detection in SE platforms is very limited, for many reasons such as lack of annotated datasets and tools specifically trained on SE interactions \citep{DBLP:journals/ese/JongelingSDS17}. The nature of gender hostility in SO has been analysed by \citet{brooke-2019-condescending}. \citet{DBLP:journals/corr/abs-2004-05589} report norm violations in SO and provide manual analysis of comments to show that offence and unfriendliness also exist in SO \citep{DBLP:journals/corr/abs-2004-05589}. Stress owing to toxicity in open source communities has been investigated by \citet{Raman_2020}. The performance of different tools to detect toxicity has been investigated by \citet{Sarker_CORR}.

In summary, in SE communities, it has been shown that the use of offensive language creates friction among members and the victims may leave SE project teams by getting demotivated \citep{Sarker_CORR}. Moreover, offensive conversation may demotivate developers and waste valuable resources like working time \citep{Raman_2020}. Therefore, addressing  and reducing conflict among SE communities is crucial. Conflict in SE platforms owing to offensive language can be minimised by sentence rephrasing, which is discussed in Section 2.2.

There have been some prior works on handling offensive language. \citet{xu2010filtering} proposed an automatic sentence-level filtering approach to semantically remove offensive contents by utilizing grammatical relations among words. Likewise, rule based Chinese profanity rephrasing has been proposed by \citet{su-etal-2017-rephrasing}. Moreover, advanced language models like GPT-3 \citep{DBLP:conf/nips/BrownMRSKDNSSAA20} and BERT \citep{devlin2018bert} also offer sequence generation models, which can also be potentially be utilised for offensive language rephrasing.

\subsection{Conflict reduction in SE platforms}
Conflict is a natural disagreement resulting from perceptions, beliefs or attitudes, that can be positive or negative \citep{Deutsch_1973}. Interpersonal conflict among software development team members negatively affects the productivity of individuals and teams \citep{gren_importance_2018}. \citet{gren_importance_2018} propose conflict management training among members of agile teams for better productivity because in agile environments, team members are more likely to be exhausted by conflict \citep{gren_links_2019}. \citet{karn_measuring_nodate} show that some forms of conflict cause more damage with increased levels of frequency and intensity of conflict in SE development teams. \citet{Gobeli_1998} analyse the causes and techniques to manage conflict in software development. Conflict due to offensive behaviour is because the community has a code of conduct which indicates politeness and mutual respect. However, behaviour exhibited by certain members in reality may not match what is expected\textsuperscript{\ref{footnote2}}. For example, offensive language is against the politeness norms. 

\section{RQ1: Prevalence of offensive language}
\label{RQ1}
Section ~\ref{RQ1:methodology} details the methodology that we employed to answer RQ1 and Section ~\ref{RQ1:Results} presents the results.

\subsection{Methodology}
\label{RQ1:methodology}
\subsubsection{Data collection}
We collected comments from four different SE communities. The datasets for GitHub, Gitter and Slack were obtained from prior work of \citet{Raman_2020}, \citet{Parra_MSR} and \citet{Chararjee_MSR}, respectively. We downloaded the publicly available SO data-dump of comments from the SO archive\footnote{\label{footnote1}\url{https://archive.org/details/stackexchange}}. We randomly chose 10\% of comments from Gitter and, 5\% of comments from SO for the month of November across the years to keep the number of comments sampled from these datasets more or less equal. The sizes of the datasets after sampling are shown in Table ~\ref{Testing Comment_Count}.

\begin{table}[h]
\centering
\caption{\label{Testing Comment_Count} Comment dataset details }
\begin{tabular}{ccc}
\toprule
\textbf{Platform} & \textbf{Comment count} & \textbf{Date range}\\    \midrule
GitHub  & 237000 & 2012-2019 \\
Gitter  & 229250 & 2014-2019 \\
Slack & 408100 & 2017-2019 \\
SO & 280400 & 2008-2019 \\
\bottomrule
\end{tabular}
\end{table}

To identify the offensive comments, we obtained the Perspective API (PAPI)\footnote{\url{https://www.perspectiveapi.com}} score and Regular Expression (Regex) status of the comments since these measures have been used by the SO Heat Detection bot\footnote{\url{https://chat.stackoverflow.com/search?q=heat&user=6294609&room=}} to detect toxicity in SO comments. PAPI offers a `toxicity' range for a text in the interval of 0 and 1, where 0 represents least toxic and 1 represents extreme toxicity. In order to locate the obfuscated offensive terms (e.g., a\$\$hole), where PAPI fails, we used a set of Regex that are used by SO bot. The complete Regex list and classification can be found online\footnote{\url{https://doi.org/10.6084/m9.figshare.14518590.v1}}. Using this approach, we can locate comments that contained misspelled or obfuscated offensive terms. The analysis and classification of comments using Regex and PAPI are detailed below.

\subsubsection{Manual classification}
Using the PAPI score and Regex check, we analysed the comments of the four communities listed in Table ~\ref{Testing Comment_Count}. We chose comments with a PAPI score of 0.7 or greater, since the comments with PAPI score beyond that threshold were found to be deleted from SO (i.e., they are truly offensive). Then we manually evaluated the comments of these platforms that had both a Regex match presence and a PAPI score >=0.7.  Subsequently, we used multi-label classification of offensive comments based on the taxonomy generated by \citet{DBLP:journals/corr/abs-2004-05589}. For this work, the taxonomy has three classes: Personal, Racial and Swearing. Any personally targeted comment like name calling, intimidating or bullying comes under the category of `Personal'. Use of offensive language to abuse one's ethnicity, skin tone or any other kind of racial discrimination is labelled as `Racial'. Use of swearing words or  abusing one's sexual orientation is labelled as `Swearing'. The first author manually labelled the above mentioned comments (i.e., those having both a Regex presence and a PAPI score>=0.7) of the four datasets into these three classes. Then we chose 100 random comments for an Inter-Rater Reliability (IRR) check from each community and obtained a Cohen's Kappa score of 0.79 among the first author and a recruited participant, showing substantial agreement on labelling \citep{McHugh_2012}. The results of classification are presented in Section 3.2.

\subsubsection{ML-based classification}
In order to build ML models to detect offensive language, we used the dataset from SO, which has been released as part of the 2020 WOAH workshop\footnote{\url{https://www.workshopononlineabuse.com}}. The dataset has been labelled by SO for research and development. There are five labels in the dataset, namely Rude or Offensive (RO), Unwelcoming or Unfriendly (UN), (Declined) RO, (Declined) UN and `None'.

The RO and UN comments are the sets of comments labelled as rude/offensive and unwelcoming/unfriendly respectively by SO moderators, after being flagged accordingly by SO community members or bot. Though the use of offensive language is milder in UN compared to RO, these comments were deleted from the site as part of moderation. The (Declined) RO and (Declined) UN are the sets of comments the moderators allowed to be in the site, by overriding the flagging by community members or a bot. For ML model creation, we used the RO comments alone from the SO dataset since UN comments are borderline cases. After  removing  duplicate entries, non-English comments and comments holding emojis alone, a total of 30787 RO comments were identified for training.

To create a model that can identify offensive and non-offensive comments, we needed a dataset that comprised comments that were offensive and non-offensive. So, we included an equal number of currently existing comments (i.e., non-offensive comments) from the publicly available SO data-dump\textsuperscript{\ref{footnote1}} to the training dataset. For validation purposes, we collected the PAPI score and Regex status of these comments to confirm that these 30787 newly added ones (non-offensive ones) are not offensive. These non-offensive comments had PAPI scores less than or equal to 0.05 and they did not contain any Regex matching terms. 

In order to develop the model with more non-offensive cases (since real-life data contains more non-offensive comments than offensive ones), we augmented the non-offensive comments using text augmentation \citep{DBLP:conf/coling/RischK18}. We implemented the text augmentation using the TextAugment library of Python \citep{marivate2020improving}, where random words in a  given statement are replaced with synonyms to produce syntactically matching statements. We included twice the amount of augmented non-offensive comments along with non-offensive comments to the ML training corpus. To retain the context of  comments that BERT needs to perform better \citep{understanding_2019}, we did not eliminate pronouns, prepositions or conjunctions from the training corpus, which is usually done in the text classification pipeline. Therefore, the final training dataset comprised of 123148 comments (30787 offensive and 92361 non-offensive).

To build ML models for offensive language detection and classification, we adopted both traditional and deep learning ML algorithms. For building conventional ML models, we chose Random Forest (RF) \citep{breiman2001random} and Support Vector Machine (SVM) \citep{vapnik1995t} as candidate models since these models are observed to be superior performers in the classification of short texts like social media comments \citep{savarimuthu2020using}. The features that we used in these model building are Regex status and sentiment features  along with TF-IDF Vectorizer \citep{ref1}. The sentiment-related features used in our work are sentiment polarity at the comment level, which can be positive, negative or neutral, and polarity score, which indicates how strongly the sentiment has been expressed in the comment. For analysing the sentiment, we used the VADER (Valence Aware Dictionary for Sentiment Reasoning) Python library \citep{DBLP:conf/icwsm/HuttoG14}. We implemented the models using Python's Scikit-learn package with default settings.

To build deep learning models, we chose BERT \textendash{}  the renowned language model developed by Google AI \citep{devlin2018bert}. BERT was chosen for the classification since it has an excellent track record of offensive language detection and classification \citep{DBLP:journals/corr/abs-2004-03705}. BERT is unsupervised and pre-trained on Wikipedia's plain text corpus. BERT has multiple variants (e.g., BERT-Base, BERT-Large-Uncased) based on the number of layers and parameters. For our classification task, we used the BERT-Base-Uncased version, which has 768 hidden layers and 110 million parameters. We implemented the classification using the  Huggingface Transformers Python package \citep{wolf-etal-2020-transformers}, with default parameters. The implementation can be found online\footnote{\url{https://github.com/jithincheriyan/Offense-detection-in-SE-communities}}.

\subsection{Results}
\label{RQ1:Results}
\subsubsection{Results of manual analysis}
\begin{figure}[!htbp]
    \setlength\abovecaptionskip{-18pt}
    \includegraphics[width=\linewidth]{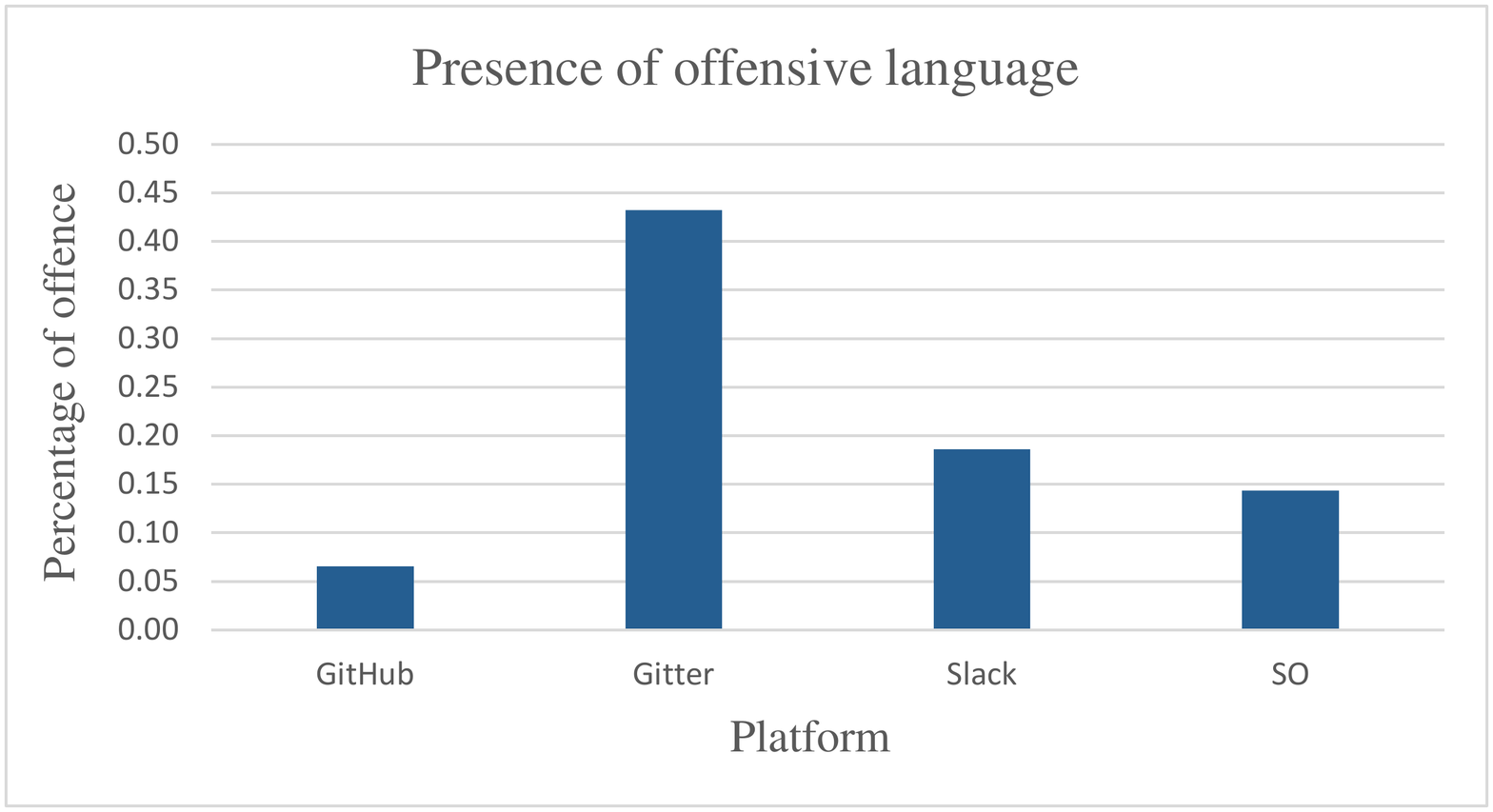}
    \caption{Total violation in four platforms}
    \setlength\belowcaptionskip{-20pt}
    \label{fig:Result1}
\end{figure}

Figure ~\ref{fig:Result1} compares the prevalence of offensive contents in the four SE platforms considered. The Gitter platform contained the most offensive comments (0.43\%) followed by Slack (0.19\%) and SO (0.14\%). GitHub comment dataset contained the least amount of offence (0.07\%). The identified offensive comments can be found online\footnote{\url{https://doi.org/10.6084/m9.figshare.14498715.v1}}.

\vspace{-33pt}

\begin{figure}[ht]
     \setlength\abovecaptionskip{-30px}
    \includegraphics[width=\linewidth]{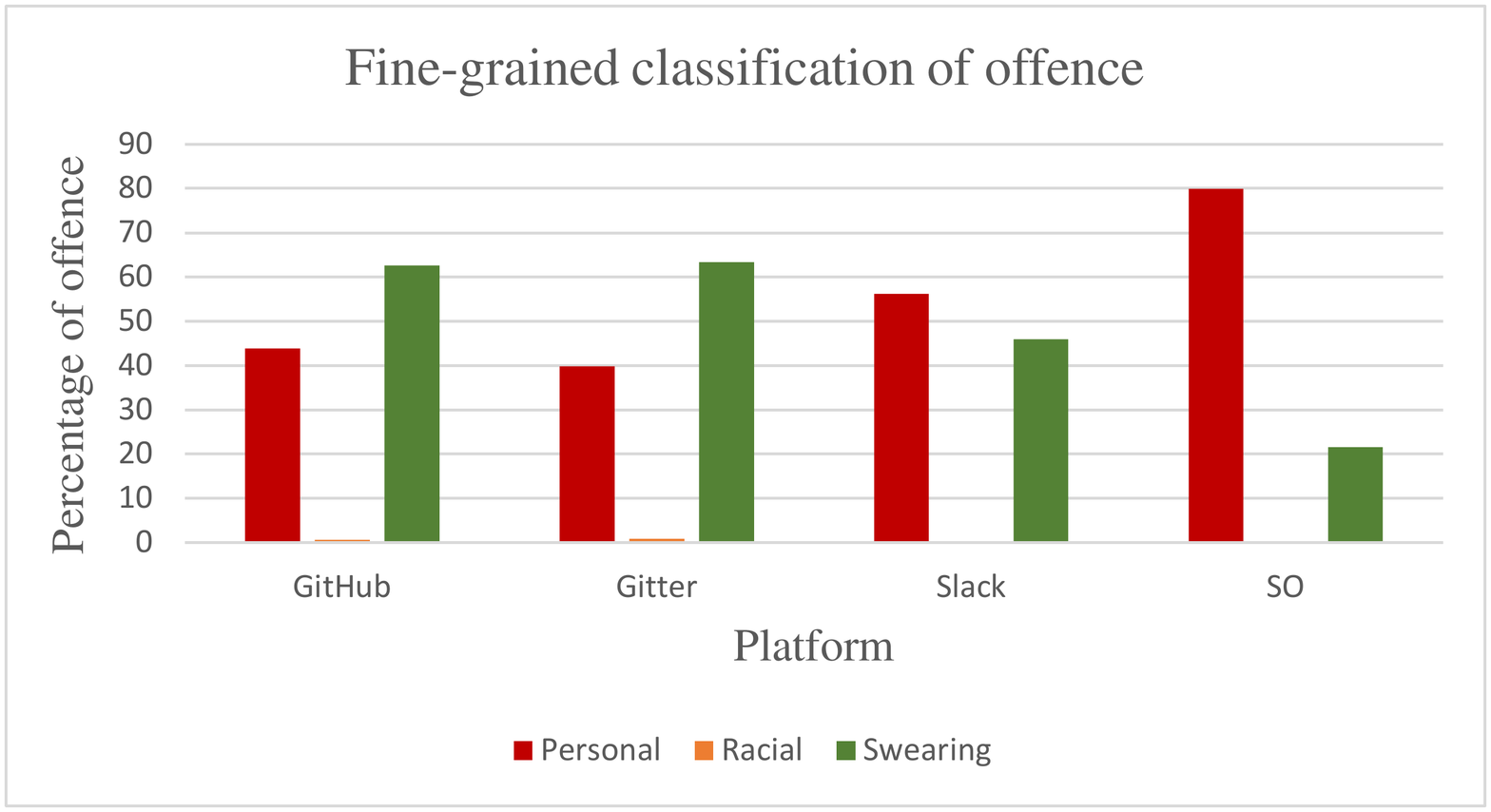}
    \setlength\belowcaptionskip{-5px}
    \caption{Fine-grained classification of offensive language in SE platforms}
    \label{fig:Result2}
\end{figure}

\vspace{-8pt}
Figure ~\ref{fig:Result2} shows the fine-grained classification of offensive language in these platforms out of the total offence present\footnote{Fine-grained results may sum up to more than 100\% owing to multi-labelling.}. In GitHub and Gitter,  Swearing is the main reason for offence (62\% and 63\% respectively), and Personal offence is less (43\% and 39\% respectively). In contrast, in SO, there is a big difference in the prevalence of Personal (79\%) and Swearing (21\%) offences. Similarly, in Slack, Personal offence (56\%) is more common than Swearing (45\%). In all these platforms, Racial offence frequency is less than 1\% and Slack had none (0\%).

\vspace{-35pt}

\vspace{-5pt}
\subsubsection{Results of ML classification}

Table ~\ref{Table: ML results} compares the performance of various ML models on offensive language detection in the four SE platforms, out of the manually collected offence. Among the models, BERT outperforms the other two substantially by offering more than 97\% detection accuracy across the SE platforms. However, SVM offered better accuracy than RF on all the platforms.

\begin{table}[ht]
\centering
\caption{\label{Table: ML results} Comparison of ML models }
\begin{tabular}{ccccc}
\toprule
\textbf{Platform} & \textbf{Manual} &\textbf{RF}&\textbf{SVM} & \textbf{BERT}\\    \midrule
GitHub & 155 & 118 (76.1\%) & 123 (79.3\%) & 151 (97.4\%) \\
Gitter & 991 & 959 (96.8\%) & 966 (97.5\%) & 983 (99.2\%) \\
Slack  & 760 & 615 (80.9\%) & 647 (85.1\%) & 756 (99.5\%) \\
SO     & 403 & 319 (79.2\%) & 371 (92.1\%) & 398 (98.8\%) \\
\bottomrule
\end{tabular}
\end{table}

\vspace{-10px}
\section{RQ2: Conflict reduction in SE communities}
\label{RQ2}
\begin{figure}[H]
    \setlength\abovecaptionskip{-15pt}
    \includegraphics[width=\linewidth]{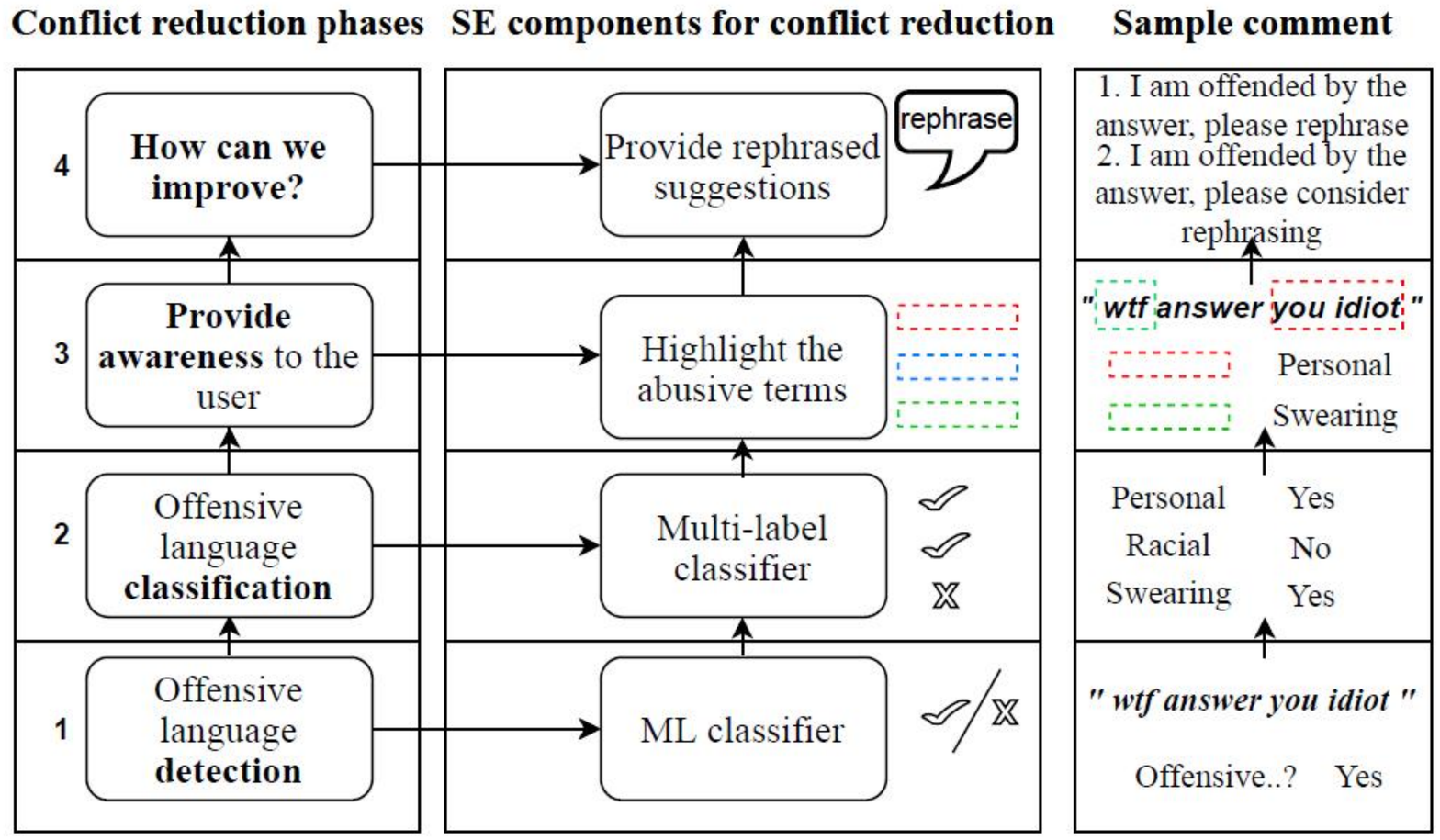}
    \caption{Conflict Reduction System framework}
    \label{fig:Conflict reduction system framework}
\end{figure}

To address RQ2, we propose CRS, a hybrid system, shown in Figure ~\ref{fig:Conflict reduction system framework} as a high-level architecture for how offensive language can be minimised in SE platforms. An offensive comment entered by a user will be processed using a four phased approach. The first phase employs a binary classifier (a comment is offensive or not) using the deep learning approach mentioned in the previous section. The second layer employs a multi-label classifier to classify the offence to a nuanced version to inform the user of nature of offence. The third level highlights the offensive contents and mentions the type of offence to the writer. Once a comment has been classified as offensive, a rule-based system comes into play to highlight the offensive terms (i.e., using dashed lines around the offensive terms), which are the reasons for a comment being labelled as offensive. This is identified using Regex checking mentioned in Section 3.

\vspace{-23pt}

\vspace{-10pt}
Even after being informed, if the writers like to express disagreement with a post, they must be given a provision for this. Therefore, the final phase of the system paraphrases the whole comment to reduce the conflict, without changing the  disagreement tone of the writer. We have developed a set of sentence rephrasing/paraphrasing tools, which provide a set of three paraphrased statements by eliminating offensive terms or substituting the offensive terms with a non-offensive one. Though the offensive terms are eliminated, we do not change the tone of the comment (e.g., negative tone) as it may infringe the rights of the user to disagree. Below we detail the paraphrasing options that we used.
 
 \paragraph{Using synonym generation} By using the milder synonyms of offensive terms identified, we are able to paraphrase offensive comments. We used the NLTK Wordnet package of Python for this purpose \citep{DBLP:journals/lre/Wagner10}. Although it paraphrases offensive comments, the overall utility of this approach is somewhat limited since the synonyms generated might not always fit the context of the input comment and might not be grammatically fitting also.
 
 \paragraph{Using a masked language model} We created a BERT-Base-Uncased model \citep{devlin2018bert}, which masks the identified offensive terms with a masking term like [MASK] so that the writer will be informed about the offensive terms. The writer may post the comment with the mask, so that he has the chance to express his disagreement without using offensive terms explicitly, which reduces friction among members. Also, from the perspective of viewers, they are free from toxic contents and conflict.
 
 \paragraph{Using a sequence-to-sequence model} We pilot-tested a sequence-to-sequence model using the Seq2Seq language model of Huggingface Transformers \citep{wolf-etal-2020-transformers}. For the model, we used the pre-trained BART-Large version from Facebook AI as the encoder and decoder \citep{lewis-etal-2020-bart}. We fine-tuned the Seq2Seq model with just 500 offensive language sentences and corresponding paraphrased statements with no abusive words. The training data can be found online\footnote{\url{https://doi.org/10.6084/m9.figshare.14518602.v1}}. Similar to the masked model mentioned above, the author can post the milder version of the offensive comment, so that viewers are less likely to be offended by seeing that. 
 
 A screenshot showing examples of above mentioned strategies for CRS can be found online\footnote{\url{ https://doi.org/10.6084/m9.figshare.14518599.v1}}. We developed a standalone system and we intend to provide a web-based interface in the future.

\section{Discussion and conclusion}
\label{Discussion}

This section discusses the implications of results presented in Sections 3 and 4. We found that the use of offensive language is prevalent in the SE platforms considered, ranging from 0.07\% to 0.43\%. Gitter has the most offence (0.43\%) because of lack of community moderation and abuse detection bots. Moreover, being a closely connected community, it is expected that the members tend to be informal to use abusive words, without the intention of hurting \citep{DBLP:conf/lrec/PamungkasBP20}. This can be the reason for substantial amount of abuse enter the Gitter discussion space. It attributes towards the need of community moderation and/or abuse detection bot to minimise offensive contents in online SE platforms.

Although abuse detection bot exists in Slack, without community moderation, many offensive contents bypass the bot by obfuscated offensive terms. This is the reason for considerable amount (0.19\%) of violation in Slack. While SO uses the Heat Detection bot and community moderation, offensive language  still exists in the site (0.14\%), which is inline with the blog of SO officials that notes that nearly 0.3\% of the total comments in SO are abusive \citep{silge_welcome_2018}. Due to the use of community moderation  and offence detection bot, the amount of violation is the least in GitHub (0.07\%). This attests the fact that human moderation along with ML techniques can do better towards defending against abuse in open SE communities. Still, not all offensive comments have been culled which points to the need for more work in this arena.

While addressing the second part of RQ1 (offence classification), we found that personally targeted abuse is more than the other two classes in two SE platforms (i.e., GitHub and Gitter). In Gitter, Swearing violations are 20\% more common than Personal abuse, and personally targeted offence is the least common among the communities considered. This is in line with our previous argument that members know each other in such a closed community, so they may use more informal swearing terms while commenting without the intention of hurting each other \citep{DBLP:conf/lrec/PamungkasBP20}.  

In Slack, Swearing comments are less frequent than personal abuse, which can be due to the functioning of the bot screening. In SO, it is observed that 79\% of the total violations are personally targeted, while the platform has the least swearing violations. This can be because of the bot, which filters all the posts holding `popular' swearing terms.
 
 Analysis of the performance of the ML models for offence detection shows that out of the manually detected comments, BERT outperforms the traditional models on all SE platforms. However, BERT has additionally detected  some comments that have neither Regex nor PAPI>=0.7. They are 1653, 4973, 7002 and 2139 in number for GitHub, Gitter, Slack and SO, respectively. However, these sets of comments were found to be unwelcoming (borderline cases) based on sampling of 50 comments in each community. The first author randomly chose 50 comments from each community and observed that, on average, 74\% of these comments are unwelcoming. This forms the basis of our future work to detect unwelcoming comments (implicit offensive language) from SE platforms. While our work shows that there is varying amount of offence in SE communities, these communities are not devoid of offence. Considering SO as an example, out of 76 million comments, 106400 comments are offensive that have not been deleted, which is significant. Our second research question aims at tackling this issue.

While addressing RQ2, it has been observed that when the recent most promising text classification models such as BERT are fine-tuned on a smaller dataset, it produces high variance on slightly different data of external datasets \citep{risch-krestel-2020-bagging}. That is, such models might not be able to perform satisfactorily on external validation like that of internal validation. To overcome that, we plan to use ensemble of models to detect and classify comments. Once offensive contents are detected and classified as discussed in the previous sections, reminding writers about the presence of the offensive terms would benefit them and the community.  When the offensive terms are highlighted, writers can have second thoughts, so that they may rethink and may abstain from posting that offence or may choose something less offensive to express the disagreement. Thereby, the writer can save their reputation points or salvage themselves from being banned from the site. The community also benefits by being free from toxic contents.  

The same strategy is followed in paraphrasing as well, since recommending milder versions of offensive terms would make the writers post less toxic contents to the discussion space and can save users' reputation in the community. Thus a win-win culture can be developed in SE communities. Our efforts here are the starting point towards the journey of effective approaches by minimizing offence in online SE communities. User evaluation and feedback will need to be pursued in the future work.

Even though offensive language elimination in online communities might be a lofty goal, partial fulfillment is possible by using ML advancements. In this work, we aimed to detect and classify offensive language in SE communities to reduce unpleasant posts by signalling to the writer about the offence in the post and by suggesting possible replacements. Using SO dataset and cutting-edge ML technologies like BERT and BART, we proposed CRS to attain the goals. A current limitation of this work is that we only consider offensive comments (swearing or profanity), not the subtler ones (unwelcoming comments). Evaluating the effectiveness of our approach by user testing is the short-term goal of this research. Overall, using this approach we believe that the conflict among community members can be reduced and the platforms can become a more welcoming place. 

\begin{acks}
I would like to acknowledge Amritha Menon for the annotation of comments as part of the Inter-Rater Reliability check.
\end{acks}

\balance
\bibliographystyle{ACM-Reference-Format}
\bibliography{Ease2021.bib}

\end{document}